\documentclass{desyproc}
\usepackage{xcolor}
\usepackage{graphicx}% Include figure files
\usepackage{bm}
\usepackage{subcaption}
\usepackage[numbers,sort&compress]{natbib}
\usepackage{upgreek}
\usepackage{wrapfig}

\begin{document}
%------------------------------------
\title{Simulation studies for the MADMAX axion direct detection experiment}

%for single authors the superscripts are optional
\author{{\slshape Jan Sch{\"u}tte-Engel$^{1}$ on behalf of the MADMAX collaboration}\\[1ex]
$^1$University of Hamburg, Hamburg, Germany }

% please enter the contribution ID for the DOI
\contribID{Lindner\_Axel}

% TO THE CONFERENCE EDITORS: 
% please update the following information      
% before sending the template to the authors
\confID{20012}  % if the conference is on Indico uncomment this line
\desyproc{DESY-PROC-2018-03}
\acronym{Patras 2018} % if you want the Acronym in the page footer uncomment this line
\doi  % if there is an online version we will register DOIs

\maketitle

\begin{abstract}
We present a general approach to solve the Maxwell-axion equations for arbitrary geometries and materials. The approach is based on the finite element method (FEM) and applied to experimental setups related to the new MADMAX (MAgnetized Disc and Mirror Axion eXperiment) project. Analytical methods are used to verify the FEM simulations. MADMAX is a dielectric haloscope which will utilize axion-photon conversion at many dielectric interfaces and probe axions in the mass range \small $m_a=40-400 $   $\upmu$eV.
\end{abstract}

\section{Introduction}
The axion \cite{PecceiQuinn,Weinberg,Wilczek} is well motivated because it solves the strong CP problem and is at the same time a dark matter candidate. MADMAX is an experiment designed to detect axions from the dark matter halo by exploiting the effect of axion-photon mixing~\cite{RaffeltStodolsky,Sikivie}. Axion-electrodynamics yields two $E$-field solutions~\cite{DishAntenna,millar_theo_foundation} when a strong B-field is applied over an interface between two materials of different refractive index, because the fields have to satisfy interface conditions. The \textit{photon-like} $E$-field is propagating away from the interface, while the \textit{axion-like} $E$-field is a non-propagating solution\footnote{Under the assumption that the $B$-field varies only slowly on the scale of the photon-like wavelength and the axions are non relativistic.}.
MADMAX will resonantly enhance the propagating waves by using many dielectric discs. An idealized 1-dimensional calculation for the $E$-field boost which MADMAX can produce already exists~\cite{millar_theo_foundation,millar_vel}. Here we present first results towards a simulation which takes into account diffraction losses and near-field effects which appear at discs of finite size.

The following text is structured as follows: In Sec. \ref{sec:MaxwellAxionEquations} the Maxwell-axion equations are introduced. After that we solve them for three different setups which are related to MADMAX. In Sec. \ref{sec:PEC} the photon-like E-fields coming from a circular perfectly electrically conducting (PEC) surface are computed. Furthermore two analytical approaches are cross checked against the 3-dimensional finite element method (FEM) simulations. In Sec. \ref{sec:DielectricDisc} the $E$-fields coming from a single dielectric disc in a strong external $B$-field are shown. The axion-electrodynamics solution for a waveguide, which is half filled with a dielectric, is computed in Sec. \ref{sec:Waveguide}.

\section{Maxwell-axion equations}\label{sec:MaxwellAxionEquations}
The Maxwell-axion equations~\cite{Sikivie} are a coupled system of partial differential equations (PDEs). To decouple the Klein-Gordon equation from the modified Maxwell equations we generalize the perturbation approach~\cite{theo_models_hoang} and expand all fields in $g_{a\gamma}$: $X(\bm{x},t) = X^{(0)}(\bm{x},t)+\sum_{i=1}^{\infty} g_{a\gamma}^i m_a^iX^{(i)}(\bm{x},t)$, with $X=\bm{E},\bm{B},\bm{J}_f,a,\rho_f$\footnote{$\bm{E}$ and $\bm{B}$ are the E and B-fields, $\bm{J}_f$ and $\rho_f$ are the free charge and current density and $a$ is the axion field.}. In the following we will only focus on the first order equations:
\begin{eqnarray}
\nabla \cdot \bm{D}^{(1)}&=&\rho_f^{(1)}+\rho_a^{(1)},
\label{Maxwell_axion_inhomogeneous_matter_gauss_e_1storder}
\\
\nabla \times \bm{H}^{(1)}-\partial_t\bm{D}^{(1)}&=&\bm{J}_f^{(1)}+\bm
J_a^{(1)},
\label{Maxwell_axion_inhomogeneous_matter_ampere_1storder}
\\
 (\partial_{\mu}\partial^{\mu}+m_a^2)a^{(1)}&=&\frac{1}{m_a}\bm{E}^{(0)}\cdot \bm{B}^{(0)},
\label{Maxwell_axion_Klein_Gordon_matter_1storder}
\end{eqnarray}
with  the axionic charge density $\rho_a^{(1)}= -\frac{1}{m_a} \bm{B}^{(0)}\cdot \nabla a^{(0)}$ and current density $\bm{J}_a^{(1)}=\frac{1}{m_a}(\bm{B}^{(0)}\partial_ta^{(0)}-\bm{E}^{(0)}\times\nabla a^{(0)})$. 
In this work we assume linear constitutive relations and no material losses $\bm{D}(\bm{x},t)=\epsilon(\bm{x}) \bm{E}(\bm{x},t)$, $\bm{H}(\bm{x},t)= \bm{B}(\bm{x},t)$, $\bm{J}_f^{(1)}=0$. Furthermore, we assume the \textit{zero velocity} limit $a^{(0)}(\bm{x},t)=a^{(0)}(t)=a_0e^{-i\omega t}$, i.e. axions are at rest and that we have only an external $B$-field (i.e. $\bm{E}^{(0)}=0$). The corresponding PDE we solve is therefore
\begin{eqnarray}
\nabla\times (\mu^{-1}\nabla \times \bm{E})-m_a^2 \epsilon\bm{E}-m_a\bm{B}^{(0)}a^{(0)}=0,
\label{vectorized_E_wave_equation_lin_media_zerovel_EBharmonic}
\end{eqnarray}
with $\omega=m_a$, the permittivity $\epsilon$ and harmonic time dependence for the $E$-field. In Eq. \eqref{vectorized_E_wave_equation_lin_media_zerovel_EBharmonic} and in all following studies the superscript $(1)$ is omitted. We use the tools COMSOL~\cite{comsol} and ELMER~\cite{elmer} to solve Eq. \eqref{vectorized_E_wave_equation_lin_media_zerovel_EBharmonic}.

\section{Perfectly electrically conducting surface}\label{sec:PEC}
\begin{figure}[h]
\centering\includegraphics[width=0.45\textwidth]{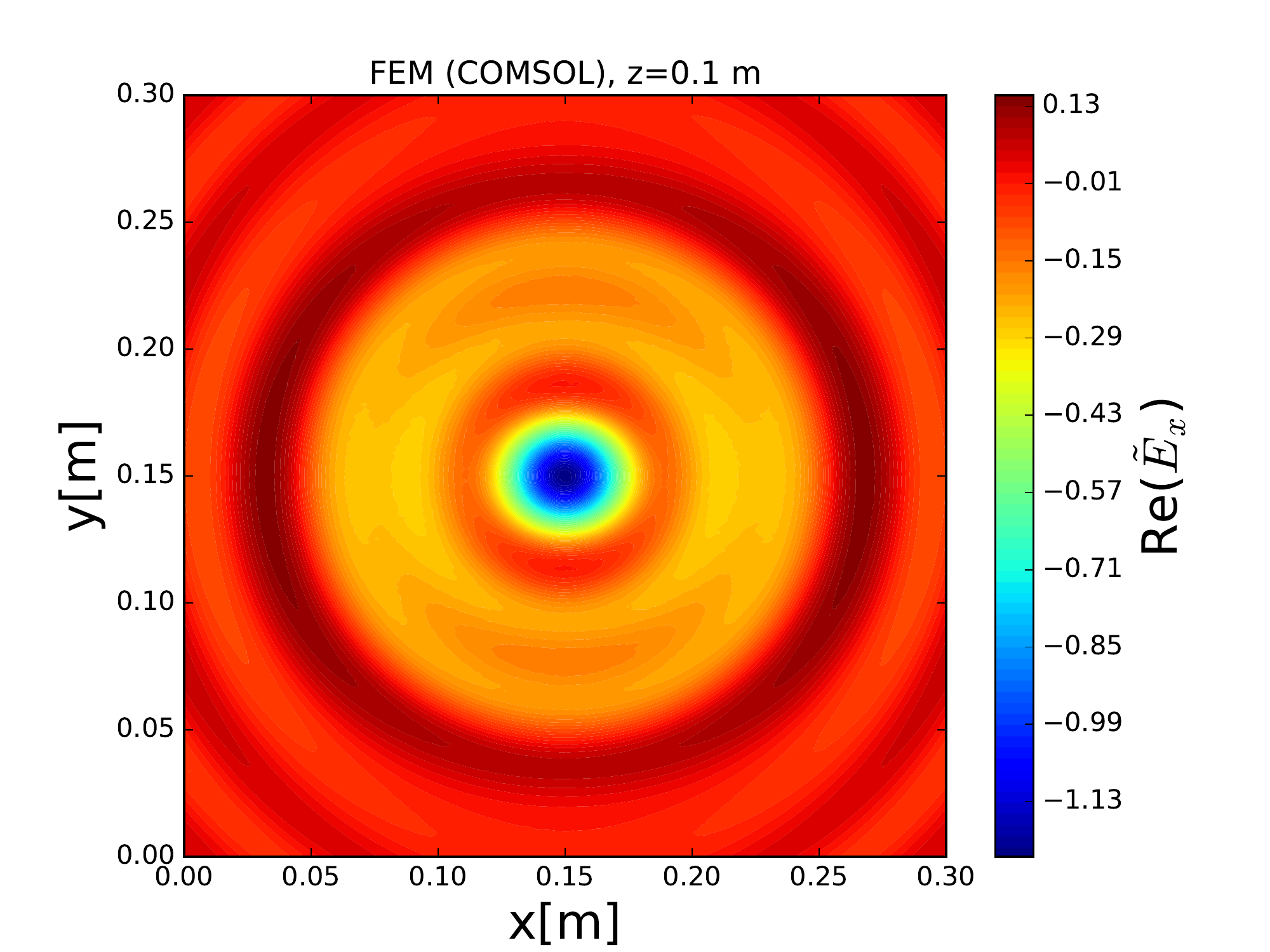}
\includegraphics[width=0.45\textwidth]{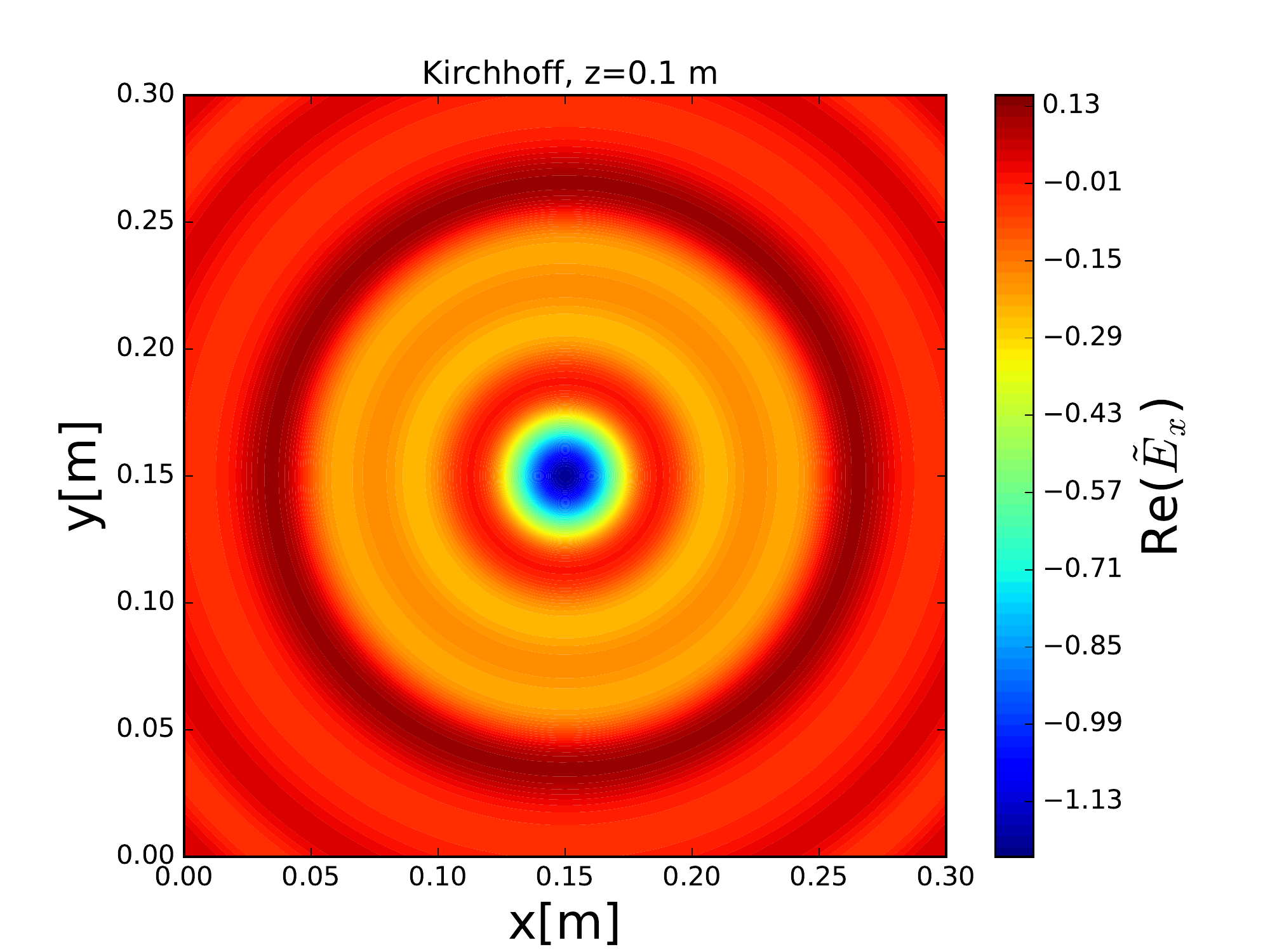}
\caption{FEM (left) and the Kirchhoff (right) results for a PEC. The tilde symbolizes that the $E$-field values are normalized to the $E$-field which is emitted by an infinite PEC surface.}\label{fig:PEC_anal_sim}
\end{figure}

In this section diffraction and near-field effects of the photon-like $E$-field from a circular PEC are investigated. The external $B$-field is assumed to be homogeneous over the PEC  $\bm{B}^{(0)}=B^{(0)}\hat{\bm{e}}_y$ as shown in Fig.~\ref{Fig:PEC}.
On the basis of a Fourier transformation~\cite{FourierOptics} we derive the following formula for the propagating $E$-field:
\begin{equation}
   \frac{E_y(r,z,t)}{E_{0}}=e^{-i\omega t}\int d\tilde{\rho}  e^{i\sqrt{\tilde{\omega}^2-\tilde{\rho}^2}\tilde{z}}J_0(\tilde{r}\tilde{\rho})J_1(\tilde{\rho}), ~~~\text{with} ~~ \tilde{\rho}=\rho R,\tilde{r}=\frac{r}{R}, \tilde{z}=\frac{z}{R},\tilde{\omega}=\omega R,
  \label{PEC_fourier}
\end{equation}  
where $J_0$ and $J_1$ are Bessel functions of the first kind and $E_0$ is the magnitude of the photon-like $E$-field at the PEC. 
To study diffraction effects we define an imaginary receiver surface at a variable distance away from the emitter surface. On the receiver surface we define the variable $\bar{U}$ as the ratio of received power over emitted power. Figure~\ref{Fig:PEC_diffraction} shows $\bar{U}$ for many frequencies and we find that for lower frequencies the diffraction loss is larger. 
A similar study for different PEC radii and a fixed frequency shows that the diffraction loss for smaller disc radii is larger. A more advanced description of the diffraction, which includes near-field effects, can be obtained by the Kirchhoff\footnote{The vector Kirchhoff formula is still not a complete description of the photon-like $E$-field, because it does not take into account boundary charges.} formula~\cite{Jackson}. In Fig.~\ref{fig:PEC_anal_sim} the Kirchhoff results are compared to the FEM results in the $xy$-plane $10\,{\rm cm}$ away from the PEC ($R=6\,{\rm cm}$ $= 2\lambda$). The overall good agreement gives us an encouraging validation of our FEM simulations.

\begin{minipage}{0.49\textwidth}
  \includegraphics[width=0.4\textwidth]{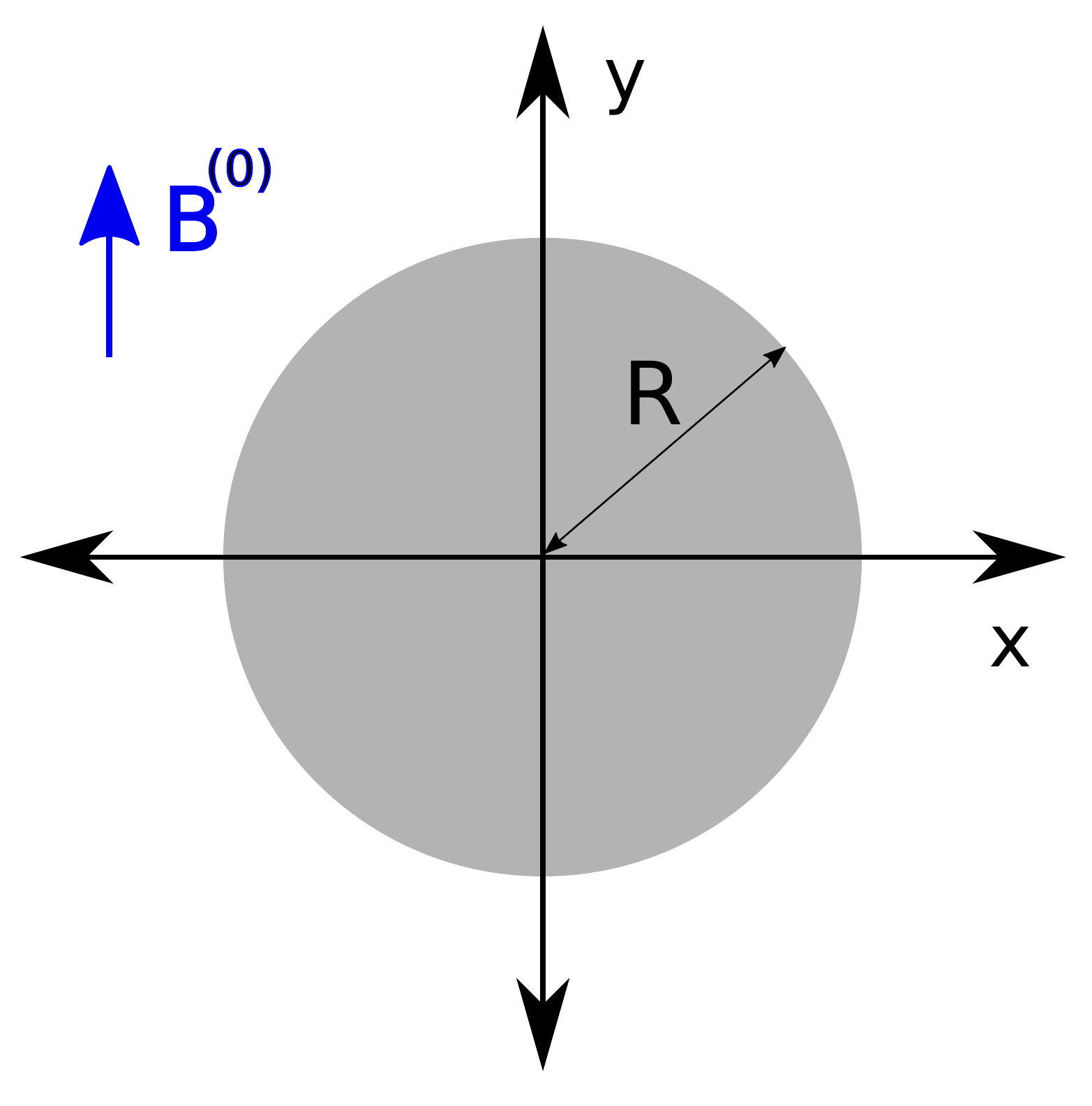}
  \includegraphics[width=0.4\textwidth]{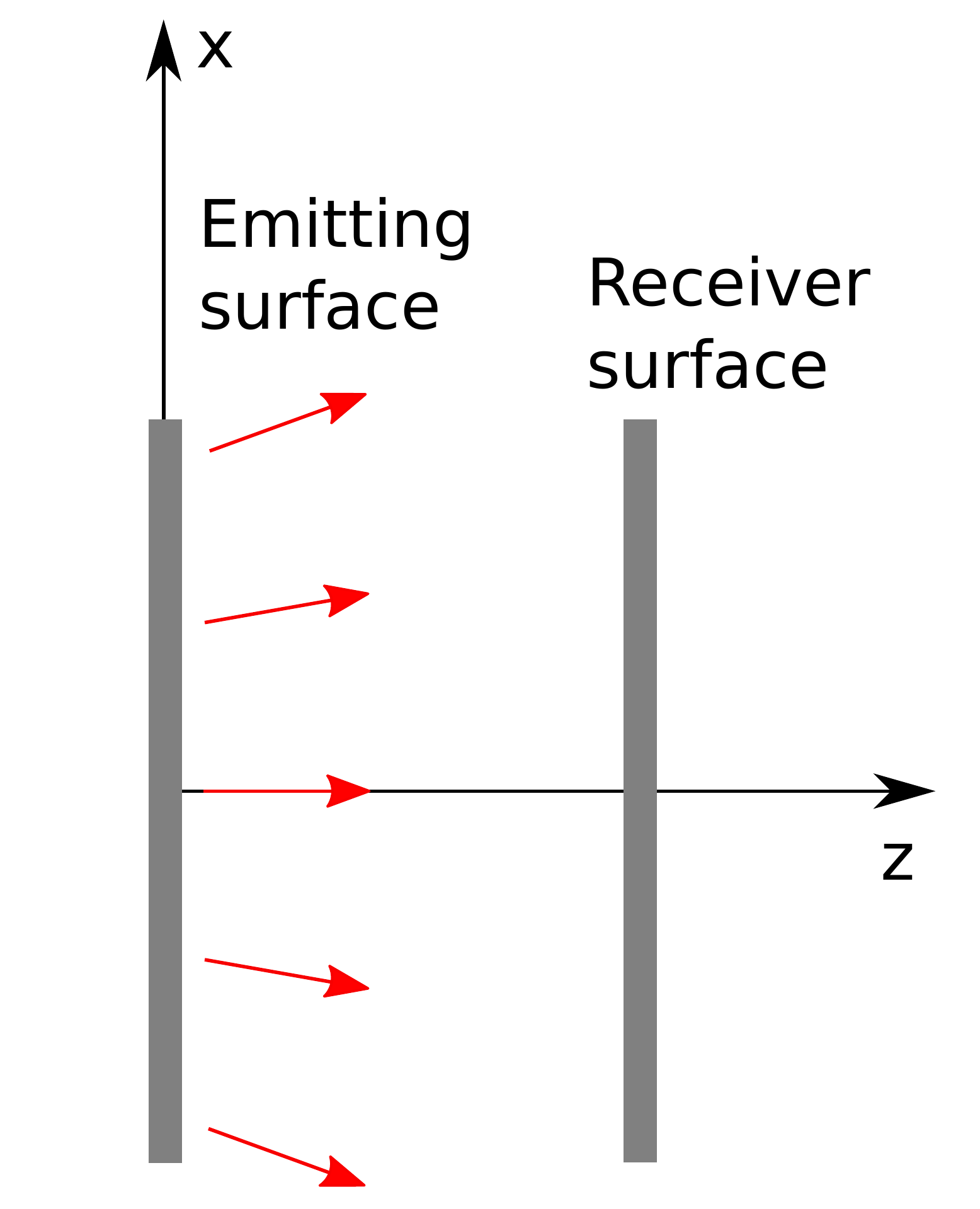}
  \captionof{figure}{PEC surface in the $xy$ and $zx$-plane.}
  \label{Fig:PEC}
\end{minipage}
\begin{minipage}{0.02\textwidth}
~~
\end{minipage}
\begin{minipage}{0.48\textwidth}
  \centering\includegraphics[width=0.7\textwidth]{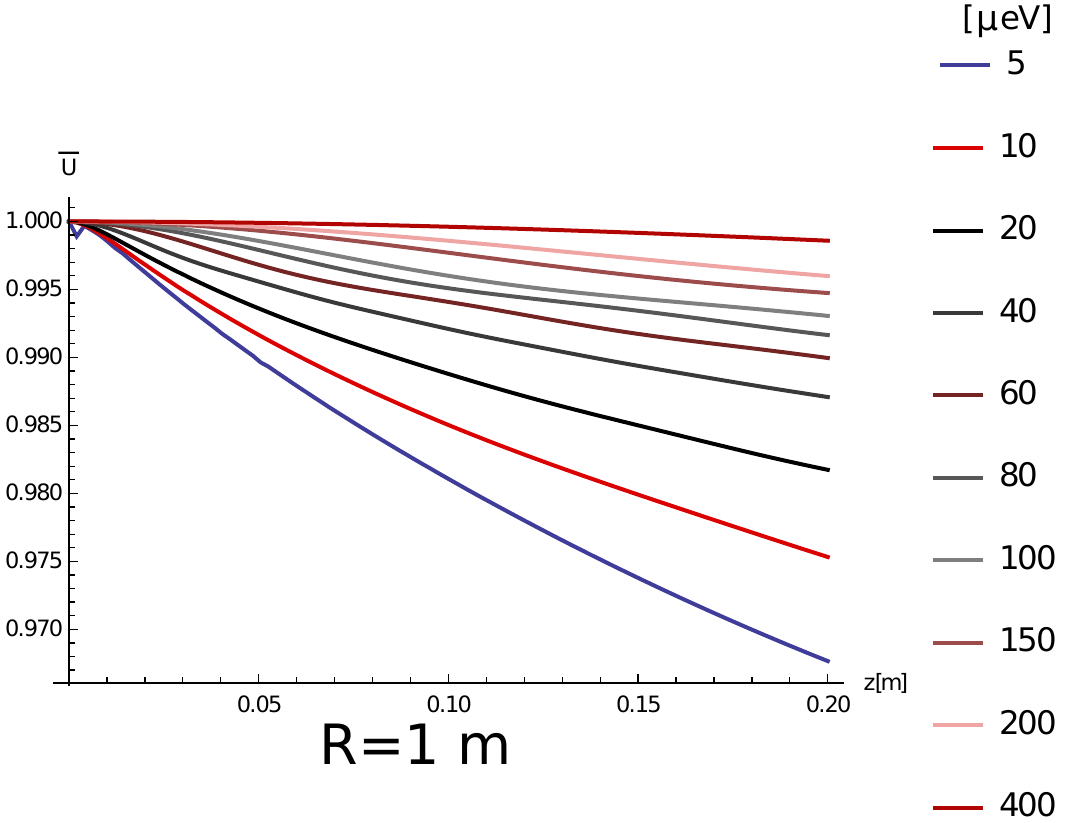}
  \captionof{figure}{Emitted power over received power for different axion masses.}
  \label{Fig:PEC_diffraction}
\end{minipage}

\section{Dielectric disc}\label{sec:DielectricDisc}
\begin{minipage}{0.45\textwidth}
  \includegraphics[width=\textwidth]{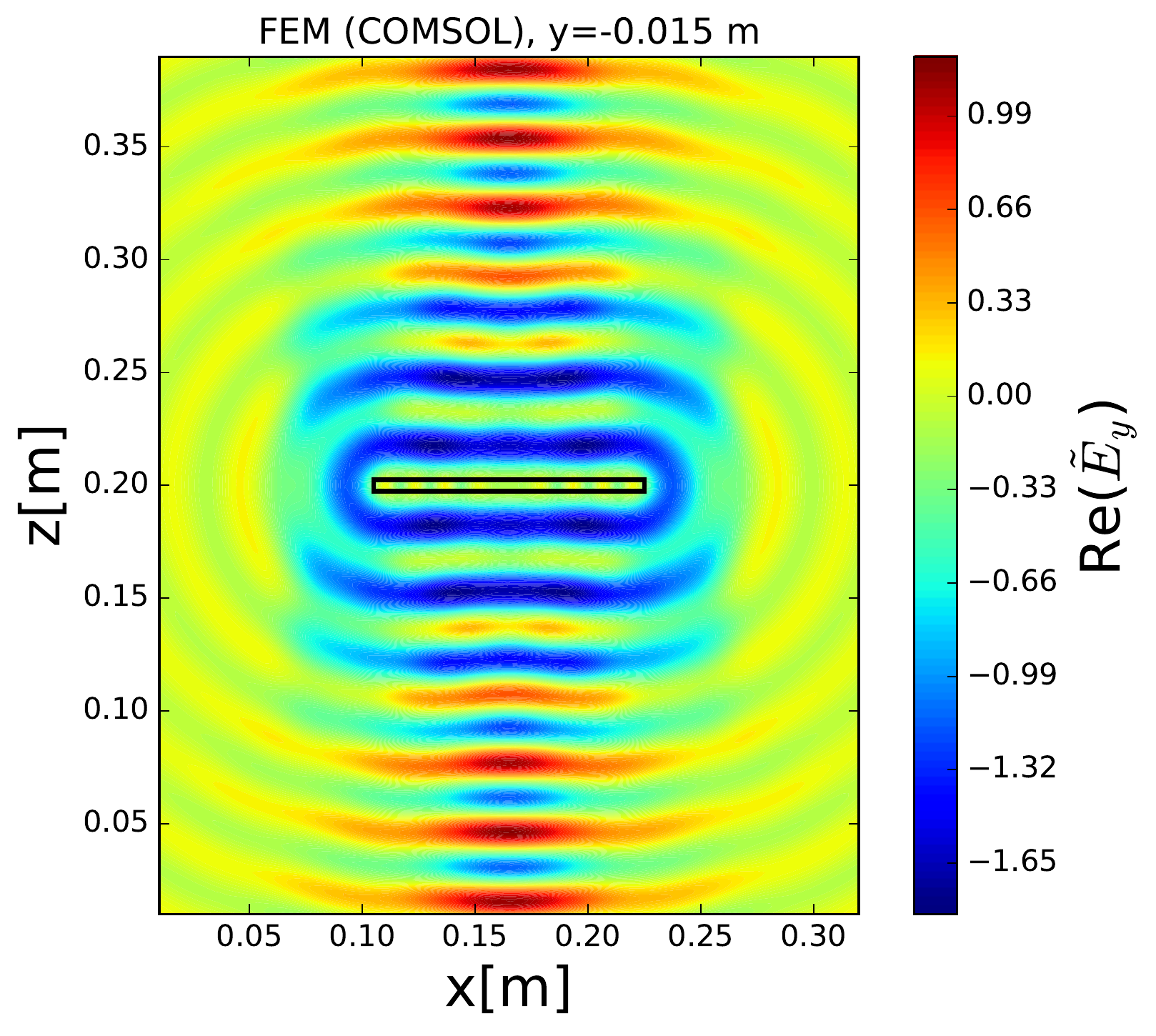}
  \captionof{figure}{Radiating dielectric disc.}
  \label{Fig:DielDisc_diffraction}
\end{minipage}
\begin{minipage}{0.02\textwidth}
~
\end{minipage}
\begin{minipage}{0.5\textwidth}
Besides a mirror the MADMAX experiment consists of many dielectric discs. In this section Eq. \eqref{vectorized_E_wave_equation_lin_media_zerovel_EBharmonic} is solved with the FEM for one circular dielectric disc of radius $R=6\,{\rm cm}$, thickness $0.5\,{\rm cm}$ and $\epsilon=9$ at a wavelength of $\lambda=3\,{\rm cm}$ (resonant case). The coordinate system is chosen as in the PEC case, just replacing the PEC with the dielectric disc (see Fig.~\ref{Fig:PEC}). The $B$-field is constant over the complete disc and drops off to the boundaries of the simulation domain. Figure~\ref{Fig:DielDisc_diffraction} shows a cross-section of the simulation domain where the disc is located at $z=0.2\,{\rm m}$. Close to the disc near-field effects are seen while far away from the disc classical diffraction effects are observed. The E-field in Fig. \ref{Fig:DielDisc_diffraction} was normalized as in the PEC case.
\end{minipage}\\

\section{Waveguide filled with dielectrics}\label{sec:Waveguide}

\begin{minipage}{0.38\textwidth}

  \centering\includegraphics[width=0.6\textwidth]{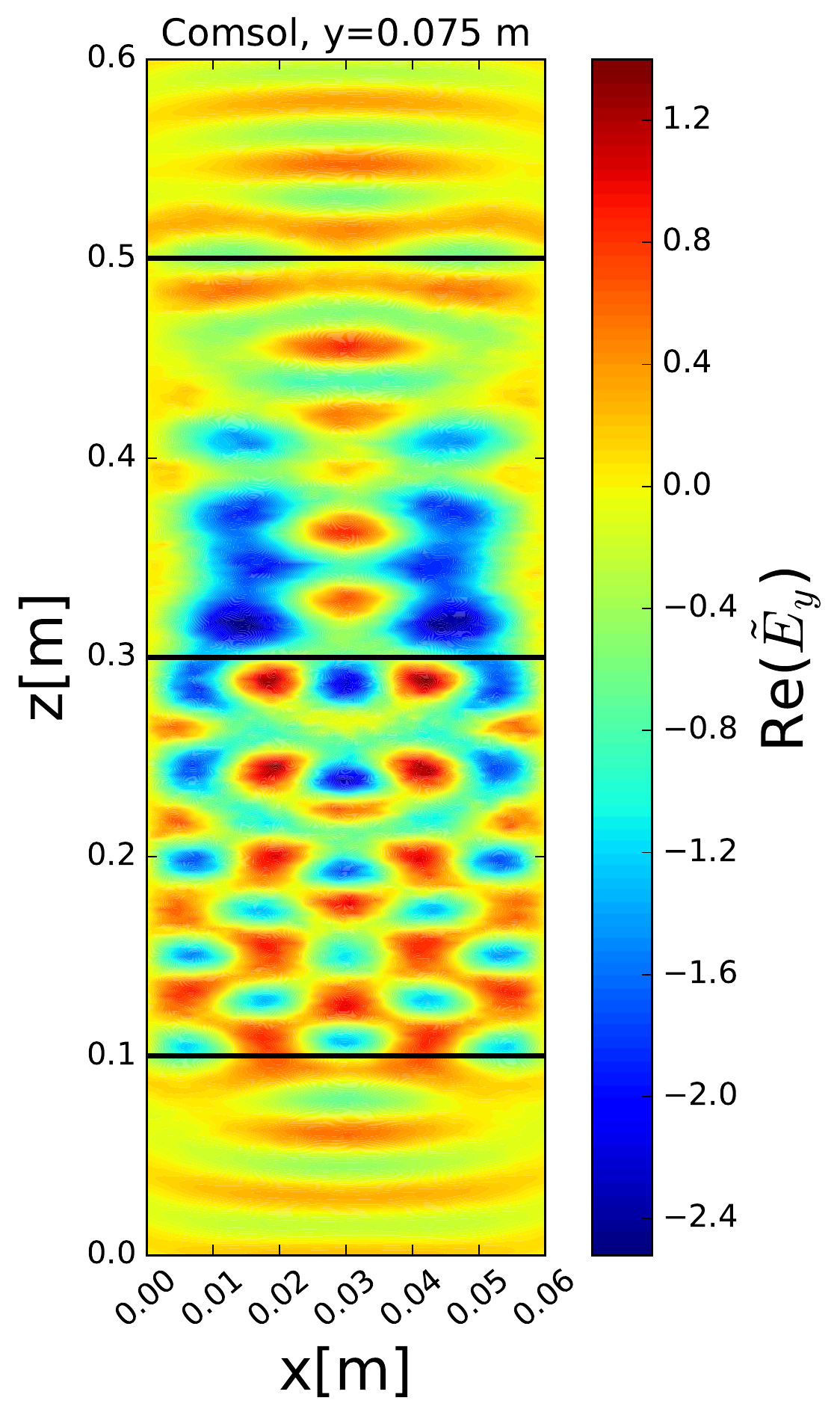}
  \captionof{figure}{$E$-field solution for a waveguide half filled with dielectric material.}
  \label{Fig:Waveguide}
\end{minipage}
\begin{minipage}{0.02\textwidth}
~~
\end{minipage}
\begin{minipage}{0.6\textwidth}
As a final example we show the simulation of a rectangular waveguide which has dimensions $2\lambda\times 5\lambda\times 20\lambda$, with $\lambda=3\,{\rm cm}$ (see Fig.~\ref{Fig:Waveguide}). The lower part of the waveguide ($0.1\,{\rm m}<z<0.3\,{\rm m}$) is filled with a dielectric ($\epsilon=2$), while the upper part of the waveguide is empty ($0.3\,{\rm m}<z<0.5\,{\rm m}$). The ranges ($z<0.1\,{\rm m}$ and $z>0.5\,{\rm m}$) are free space simulation domains.  We apply an external $B$-field which drops off in $z$-direction with a $\cos^2$ behavior to the ends of the waveguide. Inside the waveguide only specific modes are propagating which make the transition to the outside when reaching the ends of the waveguide. The tilde on the $E$-field in Fig. \ref{Fig:Waveguide} symbolizes the same normalization as in the previous Secs. 
\end{minipage}

\section{Conclusions and outlook}
We performed the first FEM simulations for MADMAX related setups. The simulations are compared to analytical calculations to validate the FEM simulations and for a better understanding of the diffraction and near-field effects, which are possible loss mechanisms in the MADMAX experiment. Our studies motivate us to investigate further loss mechanisms and to study more complicated experimental configurations, e.g. setups with more dielectric discs.

% ****************************************************************************
% BIBLIOGRAPHY AREA
% ****************************************************************************

\begin{footnotesize}
% IF YOU DO NOT USE BIBTEX, USE THE FOLLOWING SAMPLE SCHEME FOR THE REFERENCES
% ----------------------------------------------------------------------------
\begin{footnotesize}

\end{footnotesize}

\end{footnotesize}

% ****************************************************************************
% END OF BIBLIOGRAPHY AREA
% ****************************************************************************

\end{document}